\documentclass{article}
\usepackage{amssymb}
\input epsf.sty

\renewcommand{\theequation}{\thesection.\arabic{equation}}
\newcommand{\newsection}{    
\setcounter{equation}{0}
\section}
\def\appendix#1{
\addtocounter{section}{1}
\setcounter{equation}{0}
\renewcommand{\thesection}{\Alph{section}}
\section*{Appendix \thesection\protect\indent #1}
\addcontentsline{toc}{section}{Appendix \thesection\ \ \ #1}
}
\catcode`\@=11
\def\marginnote#1{}

\newcount\hour
\newcount\minute
\newtoks\amorpm
\hour=\time\divide\hour by60
\minute=\time{\multiply\hour by60 \global\advance\minute by-\hour}
\edef\standardtime{{\ifnum\hour<12 \global\amorpm={am}%
        \else\global\amorpm={pm}\advance\hour by-12 \fi
        \ifnum\hour=0 \hour=12 \fi
        \number\hour:\ifnum\minute<10 0\fi\number\minute\the\amorpm}}
\edef\militarytime{\number\hour:\ifnum\minute<10 0\fi\number\minute}

\def\draftlabel#1{{\@bsphack\if@filesw {\let\thepage\relax
      \xdef\@gtempa{\write\@auxout{\string
          \newlabel{#1}{{\@currentlabel}{\thepage}}}}}\@gtempa \if@nobreak
    \ifvmode\nobreak\fi\fi\fi\@esphack} \gdef\@eqnlabel{#1}}
    \def\@eqnlabel{}
\def\@vacuum{}
\def\draftmarginnote#1{\marginpar{\raggedright\scriptsize\tt#1}}

\def\draft{
  \oddsidemargin -.5truein
  \def\@oddfoot{\footnotesize \sl preliminary draft \hfil
    \rm\thepage\hfil\sl\today\quad\militarytime}
  \let\@evenfoot\@oddfoot \overfullrule 3pt
    \let\label=\draftlabel
    \let\marginnote=\draftmarginnote
  \def\@eqnnum{(\theequation)\rlap{\kern\marginparsep\tt\@eqnlabel}%
    \global\let\@eqnlabel\@vacuum}

  }

\newcommand{\ov}[1]{\overline{#1}}

\def\be{\begin{equation}}
\def\ee{\end{equation}}
\newcommand{\beq}{\begin{equation}}
\newcommand{\eeq}{\end{equation}}
\newcommand{\bea}{\begin{eqnarray}}
\newcommand{\eea}{\end{eqnarray}}

\newcommand{\eps}{\varepsilon}

\def\e{{\,\rm e}\,}

\renewcommand{\d}{{{\partial}}}

\begin{document}
\topmargin 0pt
\oddsidemargin 5mm
\headheight 0pt
\headsep 0pt
\topskip 9mm

{{}\hfill ITEP-TH-36/04}

\begin{center}
\vspace{26pt}
{\large \bf {AdS$_3$/CFT$_2$ on torus in the sum over geometries}}
\vspace{26pt}

{\sl L.\ Chekhov}
\footnote{email: chekhov@mi.ras.ru}
\\
\vspace{6pt}
Steklov Mathematical Institute\\ Gubkin
st.\ 8, 117966, GSP-1, Moscow, Russia\\
and\\
Institute for Theoretical and Experimental Physics\\
B.Cheremushkinskaya 25, 117259, Moscow, Russia\\
\end{center}
\vspace{20pt}
\begin{center}
{\bf Abstract}
\end{center}

\noindent
We investigate the AdS$_3$/CFT$_2$ correspondence for
the Euclidean AdS$_3$ space compactified on a solid torus with the CFT field on
the regularizing boundary surface in the bulk.
Correlation functions corresponding to the bulk theory at finite temperature
tend to the standard CFT correlation functions in the limit of removed regularization.
In both regular and ${\mathbb Z}_N$ orbifold cases, in the sum over geometries,
the two-point correlation function
for massless modes factors, up to divergent terms proportional to the volume of
the $SL(2,{\mathbb Z})/{\mathbb Z}$ group,
into the finite sum of products of the conformal--anticonformal CFT Green's functions.


\newsection{Introduction}
The AdS/CFT correspondence~\cite{Maldacena,Polyakov,Witten}
has been verified
for interacting field cases~\cite{inter,HF}
(three- and four-graviton scattering, etc.),
and it is interesting to check it also in
cases where the space--time geometry is more involved than the sphere.
Various approaches to this problem were proposed~\cite{Bonelli,Ch1,Ch}.

In~\cite{Ch1,Ch}, we considered the massless scalar
field theory on AdS$_3$ space compactified on a solid torus (toroidal handlebody).
We considered both the case of a homogeneously
compactified AdS$_3$ manifold without (topological) singularity in the
interior and the ${\mathbb Z}_N$-orbifold case.
The classical scalar field theory on the AdS$_3$ manifold
in the bulk then provided the appropriate quantum correlation functions on the boundary.

Recall that the compactification in the Euclidean case corresponds to considering
the finite temperature theory in the case of $2+1$ dimensions, so we actually calculate
correlation functions of boundary fields
for the AdS$_3$ space at finite temperature~\cite{Wit2}.
Following~\cite{Wit2}, one must take into account all possible solutions of
the Einstein gravity that have the fixed anti-de~Sitter metric at infinity.
In the AdS$_3$ case, the black-hole solutions of Hawking and
Page~\cite{HP83,Wit2} turn out~\cite{CT,MS,Ch} to be T-dual to the
case of a pure AdS$_3$ space without internal singularities;
this holds for the corresponding correlation functions as well~\cite{Ch}.
Developing these ideas, authors of~\cite{Farey} proposed that the total
CFT free energy must be developed into the sum over all possible AdS$_3$ geometries having the
same two-dimensional boundary surface. This necessarily incorporates the summation over
the modular group $SL(2,{\mathbb Z})$ factored over the parabolic subgroup ${\mathbb Z}$;
we must therefore includes the AdS geometry, the BTZ black-hole geometry, etc., but
each such geometry is equivalent to a unique AdS$_3$ geometry {\it without\/} a singularity
in the bulk, and we can as well take the sum only over such smooth geometries.

The thermal correlation functions for $1+1$-dimensional boundary theory
in the presence of the Lorentzian BTZ black hole
were obtained using the image technique in \cite{K-V};
however, our approach differs from the one used there because we do not
imply that the boundary theory is conformally invariant ad hoc. Instead, as in~\cite{Ch1,Ch},
we introduce the regularizing surface {\it inside\/} the AdS$_3$ space; this surface must be invariant
under the action of the discrete symmetry group. This approach goes in parallel with
the approach of Krasnov~\cite{Krasnov} to the 3D gravity. (In the latter,
for arbitrary genus handlebody, one must take a bounding $\eps$-surface
invariant w.r.t. the action of the corresponding Fuchsian subgroup and having the induced constant
two-dimensional curvature. This surface is determined by the height function $\eps \e^{\phi(\xi,\ov \xi)}$
with the field $\phi(\xi,\ov \xi)$ satisfying the Liouville equation.)

In Sec.~2, we recall the general structure of AdS$_3$ manifolds and introduce
the $\eps$-cone regularization in order to make the volume and the
boundary area finite in the solid torus case. In Sec.~3, we derive the Green's
function for the points located at the torus boundary in the bulk in the limit $\eps\to0$.
In Sec.~4, we take the sum over geometries for the smooth AdS$_3$ space case with the
necessary cancellations performed and demonstrate that, in the massless mode case, we attain the
holomorphic--antiholomorphic factorization of the corresponding two-point correlation
function. In Sec.~5, we investigate the corresponding sum over geometries
for the massless scalar field correlation function in the case of
${\mathbb Z}_N$-orbifold geometry and demonstrate its splitting into a finite sum of
conformal blocks. A brief discussion on massive case and perspectives is in Sec.~6.

\newsection{Geometry of AdS$_3$ manifolds}

The group $SL(2,{\mathbb C})$ of conformal transformations of the complex
plane admits the continuation to the upper half-space ${\mathbb H}_3^+$
endowed with the constant negative curvature (AdS$_3$ space).
In the Schottky uniformization picture, Riemann surfaces of higher genera
can be obtained from ${\mathbb C}$ by factoring it over a finitely
generated free-acting discrete (Fuchsian) subgroup $\Gamma\subset SL(2,{\mathbb C})$.
Therefore, we can continue the action of this subgroup to the
whole AdS$_3$ and, after factorization, obtain a three-dimensional manifold
of constant negative curvature (an AdS$_3$ manifold)
whose boundary is (topologically)
a two-dimensional Riemann surface~\cite{Manin,Sel}.

The simplest, genus one, AdS$_3$ manifold (a handlebody with the torus boundary)
can be obtained upon the identification
\be
(\xi,{\ov \xi},t)\sim (q\xi,{\ov q}{\ov \xi},|q|t),
\label{a1}
\ee
where $q=\e^{2i\pi\tau}$ is the modular parameter, $\hbox{Im\,}\tau>0$, \
$\xi,\ov \xi=x+iy,\,x-iy$ are
the coordinates on $\mathbb C$, and $t>0$ is the third coordinate in ${\mathbb H}^+_3$.

Adopting the AdS/CFT correspondence principle, we should first regularize
expressions in order to make them finite (see~\cite{Ch,Krasnov}).
For this, we set the boundary data on a two-dimensional
submanifold of the AdS$_3$ that is
invariant under the Fuchsian group action in the bulk. Such a submanifold in
the torus case is the $\eps$-{\it cone}---the set of points $(\xi,t)$ such that
$$
t=\eps|\xi|.
$$
This cone is obviously invariant w.r.t. the action of the
Fuchsian element (\ref{a1}) and becomes torus upon the identification.
The part of the ${\mathbb H}^+_3$ bounded from below by this cone becomes the interior
of the toroidal handlebody upon factorization.
Given the boundary data on this cone, we
fix the problem setting---the Laplace equation then has a unique solution
(the Dirichlet problem on a compact manifold).

Geometrically, performing the $\eps$-cone
regularization and factoring over
the group $\Gamma$ of transformations (\ref{a1}), we obtain the solid
torus on whose boundary (the two-dimensional
torus) the CFT fields dwell. The ``center'' of the torus is
a unique closed geodesic, which has the length $\log|q|$
(the image of the vertical half-line $\xi=\ov \xi=0$),
while the AdS-invariant
(proper) distance $r$ from this geodesic to the image of the $\eps$-cone is constant,
$\cosh r=1/\eps$.

Following the summation over geometries standpoint~\cite{Farey}, we must fix the two-dimensional
submanifold and consider all possible AdS$_3$ metrics with this submanifold being the boundary.
We restrict the possible class of metrics to be handlebody metrics characterized by a unique
geodesic line on the boundary homeomorphic to a unique contractible circle ($a$-cycle) in the
handlebody. The choice of the complementary $b$-cycle homeomorphic to the closed geodesic inside
the solid torus is not unique, and the freedom is exactly the Abelian (parabolic)
subgroup $\mathbb Z$ allowing adding
$a$-cycle windings to a given $b$-cycle. We therefore first re-derive the Green's function for the solid
torus in a way simpler than in~\cite{Ch} and, then, perform the summation over the geometries.
We are especially interested in the case of massless scalar field on the AdS$_3$ space. Insertions of this
field must correspond in the AdS/CFT dictionary to insertions of the $c=1$ CFT energy--momentum tensor
$\partial X\ov\partial X$ for the free scalar field $X(z,\ov z)$ of the boundary theory.
For this field, as the result, we do reconstruct (up to some divergences) the conformal block structure of the CFT
correlation function on the torus.

\newsection{Green's function in AdS$_3$}

As is well known, AdS spaces are uniform, that is, we can introduce
the interval in terms of the proper distance to the reference point,
\be
\label{dist}
ds^2=dr^2+\sinh^2r[d\theta^2+\sin^2\theta d\varphi^2].
\ee
For the integrity reasons, we present
the action of the scalar field~$\Phi$ of mass~$m$ on AdS$_3$
in coordinates (\ref{dist}),
\bea
S=\int dr\,d\theta\,d\varphi\sinh^2r\sin\theta\left\{(\d_r\Phi)^2+
\frac{(\d_\theta\Phi)^2}{\sinh^2r}+\frac{(\d_\varphi\Phi)^2}{\sinh^2r\sin\theta}+m^2\Phi^2
\right\},
\label{action1}
\eea
which, upon segregating angular degrees of freedom, results in the equation of
motion for the radial part~$\Phi(r)$ (if the total angular momentum is~$l$):
\begin{equation}\label{eom1}
\d_r\bigl(\sinh^2r\d_r\Phi(r)\bigr)-l(l+1)\Phi(r)-m^2\sinh^2r\Phi(r)=0.
\end{equation}
The Green's function for the field of mass $m$ in the bulk of the AdS$_d$ space
must therefore satisfy the equation
\begin{equation}
\label{masseq}
\sinh^{-(d-1)}r\partial_r(\sinh^{d-1}r\partial_rG(r|m))-m^2G(r|m)=\delta^{(d)}(r)
\end{equation}
with obvious conditions of decreasing at infinity. We are interested in $G(r|m)$
at the regime of large $r$ only. Then
\begin{equation}\label{Gamma}
G(r|m)|_{r\to\infty}\propto\e^{-\kappa r},\qquad \kappa^2-(d-1)\kappa-m^2=0.
\end{equation}
Recall that the mass spectrum in the AdS$_d$ is governed by the eigenvalues of
the total angular momentum operator in the
complementary sphere $S_d$, which produces the discrete mass spectrum $m^2=l(l+d-1)$,
whereas the corresponding values of $\kappa$ are integers,
\begin{equation}\label{kappa}
\kappa=d+l-1.
\end{equation}

From now on, we restrict the consideration only to the AdS$_3$ case. Then,
\begin{equation}\label{*.1}
G(r|m)=\frac{1}{4\pi}\frac{\e^{-\sqrt{1+m^2}r}}{\sinh r}=\frac{1}{4\pi}\frac{\e^{-(l+1)r}}{\sinh r}.
\end{equation}
Choosing two points, $\xi$ and $\chi$, on the complex plane and considering their images on the
$\eps$-cone, i.e., the points with the
${\mathbb H}^+_3$-coordinates $(\xi,\eps|\xi|)$ and $(\chi,\eps|\chi|)$, let
us find the distance between these two points in the bulk.
For two points in the upper half space with heights $R_1$ and $R_2$ and the distance~$d$
in the plane coordinates, the exact proper distance is
\begin{equation}\label{exactdistance}
r=\log\left[-\frac{R_2}{2R_1}-\frac{R_1}{2R_2}+
\frac12\sqrt{\left(\frac{R_2}{R_1}+\frac{R_1}{R_2}+\frac{d^2}{R_1R_2}\right)^2-4}\right]
\end{equation}
with $R_1=\eps|z|$, $R_2=\eps|w|$, and $d=|z-w|$ in our case. We can however take into account
that $d\gg |R_1-R_2|$ for any two points on the $\eps$-cone and then
\begin{equation}\label{apprdistance}
r\simeq \log\left[-\frac{R_2}{2R_1}-\frac{R_1}{2R_2}+\frac{d}{2R_1R_2}\sqrt{2R_1+2R_2+d^2}\right],
\end{equation}
and, at large distances, we merely have
\begin{equation}\label{largedistance}
r\simeq \log\left(\frac{|\xi-\chi|^2}{2\eps^2|\xi||\chi|}+O(\eps^2)\right).
\end{equation}

Passing from the Schottky uniformization to the standard complex structure on two-dimensional
torus using the exponential mapping
\begin{equation}\label{expon}
 \xi=\e^{i2\pi z},\qquad\chi=\e^{i2\pi w},
\end{equation}
and using (\ref{Gamma}),
we obtain the formula for the Green's function for two points on the $\eps$-cone
at large proper distances:
\begin{equation}\label{asympt}
G(z,w|m)=\frac{1}{4\pi}\left(\frac{|\e^{i2\pi z}-\e^{i2\pi w}|^2}{2\eps^2|\e^{i2\pi z}|\,|\e^{i2\pi w}|}\right)^{-\kappa}=
\frac{\eps^{2\kappa}}{8\pi}\left|\sin(\pi(z-w))\right|^{-2\kappa}.
\end{equation}
We keep here the scaling factor $\eps^{2\kappa}$, which will be removed only at the very end
of calculations, after the summation over geometries. The reason for it will be clarified in the succeeding
section. The function in (\ref{asympt})
is periodic under shifts $z\to z+m$ and $w\to w+k$ for $k,m\in{\mathbb Z}$.
In order to obtain the Green's function for the solid torus boundary, we must now take the sum over all
images~$z_n$ of the point $z$, \ $z_n=z+n\tau$. Thus, the final answer for the Green's function
on the toroidal handlebody boundary is
\begin{equation}\label{as-torus}
G_{\mathrm{torus}}(z,w|m)=\frac{\eps^{2\kappa}}{8\pi}\sum_{n=-\infty}^{\infty}
\left|\sin(\pi(z-w+n\tau))\right|^{-2\kappa}.
\end{equation}

This formula (up to scaling factors) exactly reproduces the answer
obtained in~\cite{Ch} in the massless case ($\kappa=2$) and the answer
in~\cite{Kach} for the general massive field.

\newsection{Summing over geometries}

The most difficult problem when evaluating the sum over geometries is the choice of the proper
summation measure. For this, different proposals had been made~\cite{Farey,Kach}.
Using ideas of~\cite{Farey}, we assume that
one must consider the {\it same\/} two-dimensional boundary surface, the proper continuation to the
AdS$_3$ being then completely determined by the choice of (contractible) $a$-cycles; in the torus case,
we have just one $a$-cycle determined by two relatively prime integers $(c,d)$ with the
identification $(c,d)\propto(-c,-d)$. The proper
modular transformation from the $SL(2,\mathbb Z)/{\mathbb Z}$ is then
\begin{equation}\label{trans}
 \tau\to\tau'\equiv\frac{a\tau +b}{c\tau +d},
 \qquad z\to z'\equiv\frac{z}{c\tau+d}, \qquad w\to w'\equiv\frac{w}{c\tau+d}, \qquad\left|{a\ b\atop c\ d}\right|=1,
\end{equation}
where the pair $(a,b)$ must be taken modulo the parabolic group transformation $(a,b)\to (a+c,b+d)$. This is because the
choice of the contractible cycle is completely determined by the pair $(c,d)$ as the image
of the straight line along the vector $c\tau+d$ upon the identification $z\simeq z+1\simeq z+\tau$.
But the choice of the complementary
$b$-cycle (the image of the straight line along the vector $a\tau+b$) is fixed only up to the freedom of adding the
$a$-cycle vector. The condition of the unit determinant in (\ref{trans}) then just expresses that these two
cycles have exactly one intersection point at the torus.

We take the summation measure to be just the {\it hyperbolic volume\/} of the toroidal handlebody bounded by the
torus in the AdS space with the $a$-cycle selected. A Graham and Lee theorem~\cite{GL} claims that choosing a
conformal metric on the boundary of a ball, the smooth AdS continuation of the metric inside the ball is unique.
The AdS metric inside a handlebody is then completely determined by the conformal metric on the
boundary Riemann surface
together with the choice of a set of contractible $a$-cycles. In order to calculate both the corresponding metric
and the volume, we use the scheme depicted in Fig.~1.

In the geometry of Fig.~1, we consider the action of the Schottky group on the complex plane ${\mathbb C}$ with the
modular parameter $\tau_{(c,d)}=\tau'\equiv\frac{a\tau+b}{c\tau+d}$. It acts by identifying the circle $|z|=1$ and
the circle $|z|=\hbox{Im\,}\tau_{(c,d)}$. Note that for the matrix $\left({a\ b\atop c\ d}\right)$ with the unit
determinant,
\begin{equation}\label{gg1}
\hbox{Im\,}\tau_{(c,d)}=\frac{\hbox{Im\,}\tau_{(0,1)}}{|c\tau+d|^2}.
\end{equation}
Next, we continue this identification using the formula (\ref{a1}) into the whole AdS$_3$.
The natural coordinates in Fig.~1 are the spherical coordinates $(\rho,\eps,\varphi)$, where
$\rho=\frac12\log(\xi\ov\xi+t^2)$ is the logarithm of the Euclidean distance to the origin and $\eps$ and
$\varphi$ are the corresponding angles; the interval (\ref{dist}) in this coordinates is
\begin{equation}\label{gg12}
ds^2=\frac{d\rho^2+d\eps^2+\cos^2\eps d\varphi^2}{\sin^2\eps}.
\end{equation}
We take such the bounding
surface for which the proper AdS$_3$
{\it circumference\/} $l^{(c,d)}_a$ of the $a$-circle is exactly proportional to the length of the
corresponding geodesic in the plane metric on ${\mathbb C}$, i.e., we set
\begin{equation}\label{gg2}
l^{(c,d)}_a=2\pi\frac{\cos\eps_{(c,d)}}{\sin\eps_{(c,d)}} =|c\tau+d|l^{(0,1)}_a=|c\tau+d|
2\pi\frac{\cos\eps_{(0,1)}}{\sin\eps_{(0,1)}},
\end{equation}
that is
\begin{equation}\label{gg21}
\frac{\cos\eps_{(c,d)}}{\sin\eps_{(c,d)}} =|c\tau+d|\frac{\cos\eps_{(0,1)}}{\sin\eps_{(0,1)}}.
\end{equation}
This choice is justified by that it leads to the proper scaling behavior of Green's functions (see below).
It then follows from the hyperbolic geometry that
\begin{equation}\label{gg3}
l^{(c,d)}_b=\frac{\hbox{Im\,}\tau_{(c,d)}}{\sin\eps_{(c,d)}} =\frac{\hbox{Im\,}\tau_{(0,1)}}{|c\tau+d|^2\sin\eps_{(c,d)}}.
\end{equation}

\vskip10pt

\centerline{\epsfysize9cm\epsffile{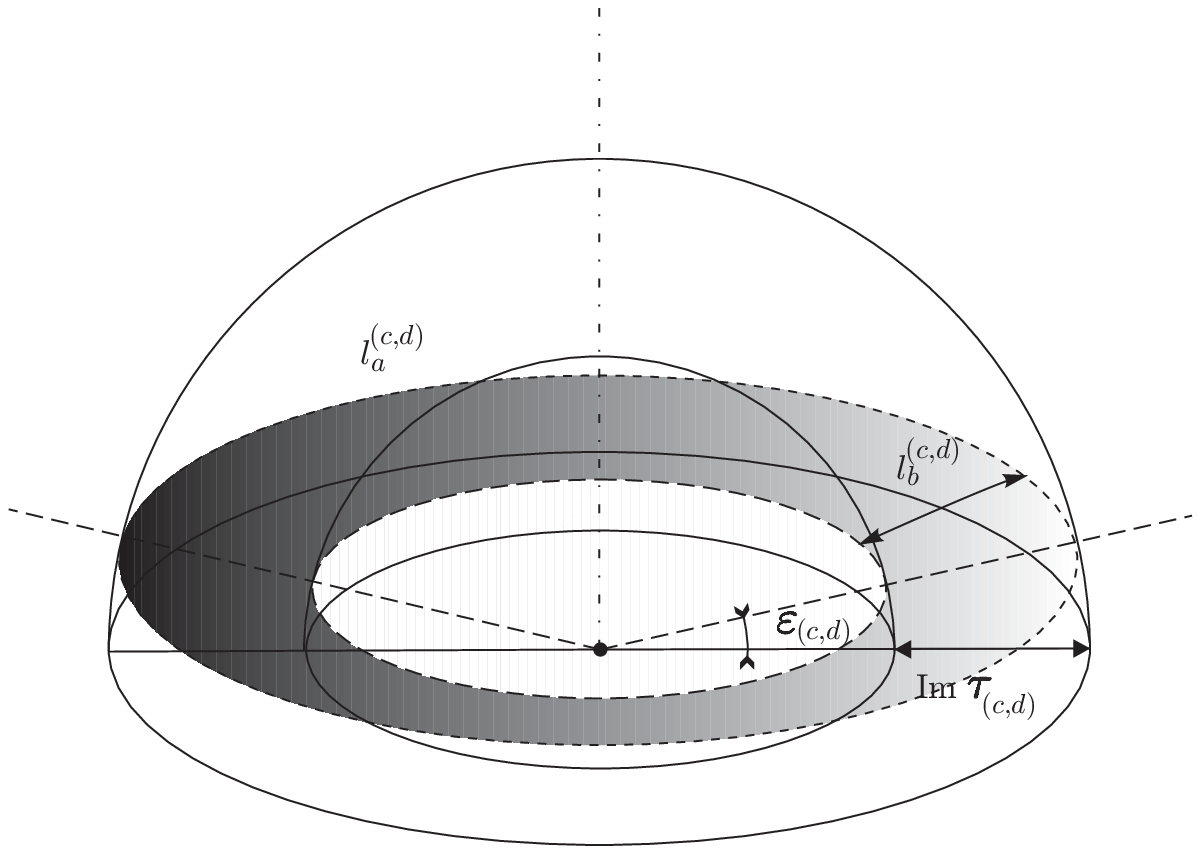}}

\vskip10pt

\noindent
{\bf Fig.~1}. The regularizing surface for the torus. Here $l^{(c,d)}_a$ is the proper (AdS) circumference of
the regularizing cone (which is actually a cylinder in the proper distance geometry) and $l^{(c,d)}_b$ is the
proper (AdS) distance between identified circles on the cone.

\vskip10pt

For the $\eps$-cone, the volume of the toroidal handlebody bounded by a torus of the induced area
$S_{(c,d)}$ lying at the distance $\hbox{arcsinh\,}\eps$ from a unique closed geodesic is
\begin{equation}\label{gg4}
V_{(c,d)}=\frac12 S_{(c,d)}\cos\eps_{(c,d)}=\frac12 l^{(c,d)}_al^{(c,d)}_b\cos\eps_{(c,d)},
\end{equation}
and using formulas (\ref{gg2})--(\ref{gg3}), we obtain the simple relation:
\begin{equation}\label{gg5}
V_{(c,d)}=V_{(0,1)},
\end{equation}
i.e., under our prescription, the hyperbolic volume of the regularizing manifold becomes exactly modular invariant,
and we have the {\it same\/} weight factors standing by
the Green's functions of the form (\ref{as-torus}). Then, due to (\ref{gg21}),
the factor $\eps_{(c,d)}^{2\kappa}$ in (\ref{asympt}) becomes in the limit $\eps\to 0$
just the proper
scaling factor $|c\tau+d|^{-2\kappa}\eps^{2\kappa}_{(0,1)}/2$
standing by the relevant Green's function: upon the transformation (\ref{trans}), we have
\begin{equation}\label{trans-Green}
G_{\mathrm{torus}}(z,w|m)\to
G_{\mathrm{torus}}^{(c,d)}(z-w|m)\equiv
\sum_{n=-\infty}^{\infty}
\left|(c\tau+d)\sin(\pi(z'-w'+n\tau'))\right|^{-2\kappa},
\end{equation}
where we have eventually omitted the irrelevant overall factor $\eps^{2\kappa}_{(0,1)}/(8\pi)$,
and it only remains to take the sum over all $(c,d)$-pairs.

Because the small-$z$ behavior of the function (\ref{trans-Green}) is obviously
\begin{equation}\label{Green-as}
\left.G_{\mathrm{torus}}^{(c,d)}(z|m)\right|_{z\to0}\simeq \frac{1}{|z|^{2\kappa}},
\end{equation}
and it is independent on the $(c,d)$-pair choice, in order to obtain the finite
answer when performing the summation over $(c,d)$-pairs, we must subtract from each (except one) term
of this sum the corresponding (doubly periodically continued) function
\begin{equation}\label{reg-F}
  F(z|\kappa)=\sum_{n,k\in {\mathbb Z}}\frac{1}{|z+n\tau+k|^{2\kappa}}.
\end{equation}

The thus regularized sum becomes especially instructive in the
{\it massless case\/} $\kappa=2$. Using the representation for the function $\sin^{-2}(\pi z)$,
\begin{equation}\label{sin2}
\frac{1}{\sin^2(\pi z)}=\sum_{k=-\infty}^\infty\frac1{\pi^2(z+k)^2},
\end{equation}
we can present the sum over geometries in the following form (omitting irrelevant $\pi$-factors):
\bea
&{}&\sum_{(c,d)}{}^{\prime\,\mathrm{(reg)}}G_{\mathrm{torus}}^{(c,d)}(z|m)|_{\kappa=2}=
\nonumber
\\
&=&F(z|1)+\sum_{(c,d)}{}'\left(G_{\mathrm{torus}}^{(c,d)}(z|m)|_{\kappa=2}-F(z|1)\right)
\nonumber
\\
&=&F(z|1)+
\nonumber
\\
&+&\sum_{(c,d)}{}'\left(\sum_{k_1,k_2,n\in {\mathbb Z}}
\frac{1}{(z+k_1(c\tau+d)+n(a\tau+b))^2(\ov z+k_2(c\ov\tau+d)+n(a\ov\tau+b))^2}-F(z|1)\right)
\label{sumcd}
\eea
(the primed sum denotes that we must count pairs $(c,d)$ and $(-c,-d)$ just once).
In this sum, we must take for each pair $(c,d)$ a single complementary pair $(a,b)$ such that
the determinant of the matrix $\left({a\ b\atop c\ d}\right)$ is the unity.

Collecting the coefficients standing by $\tau$ and unit factors in (\ref{sumcd}), we obtain
\bea
&{}&\sum_{(c,d)}{}'\left(\sum_{k_1,k_2,n\in {\mathbb Z}}
\frac{1}{(z+(k_1c+na)\tau+(k_1d+nb))^2(\ov z+(k_2c+na)\ov\tau+(k_2d+nb))^2}-F(z|1)\right)=
\nonumber
\\
&=&\sum_{(c,d)}{}'\left(\sum_{k_1,k_2,n\in {\mathbb Z}}
\frac{1}{(z+s_1\tau+t_1)^2(\ov z+s_2\ov\tau+t_2)^2}-\sum_{s,t\in{\mathbb Z}}\frac{1}{|z+s\tau+t|^4}\right),
\label{sumcd2}
\eea
where $s_{1,2}$ and $t_{1,2}$ are solutions of the system of equations
\bea
na+k_1c&=&s_1,
\nonumber
\\
na+k_2c&=&s_2,
\nonumber
\\
nb+k_1d&=&t_1,
\nonumber
\\
nb+k_2d&=&t_2.
\nonumber
\eea
Subtracting the second equation from the first and the fourth from the third, we have
\bea
\Delta k\cdot c=\Delta s;\qquad \Delta k\cdot d=\Delta t,
\label{sumcd3}
\eea
where $\Delta k\equiv k_1-k_2$, $\Delta s\equiv s_1-s_2$, and $\Delta t\equiv t_1-t_2$.
Because GCD$(c,d)=1$ and we identify $(c,d)\simeq(-c,-d)$, equations (\ref{sumcd3}) admit a unique solution
$$
\Delta k=\pm\hbox{GCD}(\Delta s,\Delta t),\qquad c=\pm\Delta s/\Delta k, \qquad d=\pm\Delta t/\Delta k
$$
unless $\Delta s=\Delta t=0$.
(Recall that $c$ and $d$ may vanish, but not simultaneously, and if $c$=0, then $d=\pm1$ and
vice versa.)
Given $c$ and $d$ (and, correspondingly, $a$ and $b$), the solution of, say, the matrix equation
w.r.t. $n$ and $k_1$,
$$
(n,k_1)\cdot\left({a\ b\atop c\ d}\right)=(s_1,t_1),
$$
exists and is unique in $\mathbb Z$ because the determinant of the matrix $\left({a\ b\atop c\ d}\right)$ is the unity.
The contribution at $\Delta s=\Delta t=0$ implies $\Delta k=0$, and this contribution is then exactly cancelled
by the contribution from the function $F(z|1)$ (\ref{reg-F}).

We therefore obtain that the regularized sum in (\ref{sumcd}),
\be
\sum_{(c,d)}{}^{\prime\,\mathrm{(reg)}}G_{\mathrm{torus}}^{(c,d)}(z|m)|_{m=0}=
\sum_{{s_1,t_1\atop s_2,t_2}\in {\mathbb Z}}
\frac{1}{(z+s_1\tau+t_1)^2(\ov z+s_2\ov\tau+t_2)^2}=\wp(z|\tau)\ov{\wp(z|\tau)},
\label{sumcd4}
\ee
becomes just the squared modulus of the Weierstrass $\wp$-function
$$
\wp(z|\tau)=\sum_{s,t\in{\mathbb Z}}\frac{1}{(z+s\tau+t)^2},
$$
We may expect that this term provides the relevant contribution to the CFT
correlation function in the sum over the AdS$_3$ metrics because it exhibits the structure of CFT
conformal blocks. Say, the function (\ref{sumcd4}) exactly coincides with the correlation function
$\left\langle\partial X(z,\ov z)\ov\partial X(z,\ov z)\partial X(0,0)\ov\partial X(0,0)\right\rangle$
for the $c=1$ free scalar field $X(z,\ov z)$ on torus.

\newsection{Summing over geometries in ${\mathbb Z}_N$-orbifold case}

Using our technique, we can find a sum over geometries for the
correlation functions of the massless field also in the case where
a conical singularity with solid angle
$2\pi/N$ is located at the closed geodesic line (the vertical axis in Fig.~1) of the toroidal handlebody.
We can resolve this singularity by considering the $N$-sheet covering of the corresponding space;
this covering is just the ``old''
toroidal handlebody. This corresponds to imposing the following double periodic conditions on
admitted functions of the scaled variable
\be
\tilde z\equiv z/N
\label{tilde-z}
\ee
in the torus case:
\be
f(\tilde z)=f(\tilde z+1),\quad f(\tilde z)= f(\tilde z+\tau).
\label{i5}
\ee
The corresponding correlation functions in the massless case are~\cite{Ch}
\begin{equation}\label{as-torus-orbi}
G_{\mathrm{torus}}^{(N)}(z|0)=\sum_{n=-\infty}^{\infty}\sum_{p=1}^N
\left.\left|\frac{1}{\pi}\sin(\pi(z+p+n\tau)/N)\right|^{-2\kappa}\right|_{\kappa=2},
\end{equation}
and in the sum over geometries, we must take~\cite{Kach} (cf.~(\ref{sumcd}))
\be
\sum_{(c,d)}{}^{\prime\,\mathrm{(reg)}}G_{\mathrm{torus}}^{(c,d);N}(z|0)=
F(z|1)+\sum_{(c,d)}{}'\left(G_{\mathrm{torus}}^{(c,d);N}(z|0)-F(z|1)\right),
\label{sumcd-orbi}
\ee
where
\begin{equation}\label{orbi-torus}
G_{\mathrm{torus}}^{(c,d);N}(z|m)\equiv
\sum_{n=-\infty}^{\infty}\sum_{p=1}^N
\left|\frac{1}{\pi}(c\tau+d)\sin(\pi(z'+p+n\tau')/N)\right|^{-2\kappa},
\end{equation}
with $z'$, $\tau'$ from (\ref{trans}). The function $F(z|\kappa)$ is given by the same
formula (\ref{reg-F}). Using the same trick as in the smooth torus case, we represent the
expression in (\ref{sumcd-orbi}) as the series
\bea
&{}&N^4F(z|1)+N^4\sum_{(c,d)}{}'\left(\sum_{k_1,k_2,n\in {\mathbb Z}}\sum_{p=1}^N\times\right.
\nonumber
\\
&{}&\times
\left.\frac{1}{(z+(Nk_1+p)(c\tau+d)+n(a\tau+b))^2(\ov z+(Nk_2+p)(c\ov\tau+d)+n(a\ov\tau+b))^2}-F(z|1)\right)=
\nonumber
\\
&=&N^4F(z|1)+N^4\sum_{(c,d)}{}'\left(\sum_{k_1,k_2,n\in {\mathbb Z}}\sum_{p=1}^N
\frac{1}{(z+s_1\tau+t_1)^2(\ov z+s_2\ov\tau+t_2)^2}-\sum_{s,t\in{\mathbb Z}}\frac{1}{|z+s\tau+t|^4}\right),
\label{orbi-t2}
\eea
where $s_{1,2}$ and $t_{1,2}$ are solutions of the system of equations
\bea
na+(Nk_1+p)c&=&s_1,
\nonumber
\\
na+(Nk_2+p)c&=&s_2,
\nonumber
\\
nb+(Nk_1+p)d&=&t_1,
\nonumber
\\
nb+(Nk_2+p)d&=&t_2,
\nonumber
\eea
which again has a unique solution if $\Delta s=Nx$, $\Delta t=Ny$ and $x,y\in {\mathbb Z}$
do not vanish simultaneously;\footnote{We conveniently express this condition as $x^2+y^2>0$.} \
the contribution when $x=y=0$ is again exactly cancelled by the function $F(z|1)$.

Eventually, we have
\bea
&{}&\sum_{(c,d)}{}^{\prime\,\mathrm{(reg)}}G_{\mathrm{torus}}^{(c,d);N}(z|0)=
\nonumber
\\
&=&\sum_{s,t,x,y\in{\mathbb Z}}
\frac{N^4}{(z+s\tau+t)^2(\ov z+(s+Nx)\ov\tau+(t+Ny))^2}=
\nonumber
\\
&=&\sum_{s_1,t_1,s_2,t_2\in{\mathbb Z}}\sum_{p=1}^N\sum_{q=1}^N
\frac{N^4}{(z+p\tau+q+Ns_1\tau+Nt_1)^2(\ov z+p\ov \tau+q+Ns_2\ov\tau+Nt_2)^2}=
\nonumber
\\
&=&\sum_{p=1}^N\sum_{q=1}^N\left|\wp_{\left[\frac pN\right]\,\left[\frac qN\right]}(\tilde z|\tau)\right|^2,
\label{orbi-t3}
\eea
where
$$
\wp_{\left[\frac pN\right]\,\left[\frac qN\right]}(\tilde z|\tau)\equiv
\sum_{s,t\in{\mathbb Z}}\frac{1}{\left(\tilde z+\frac pN\tau+\frac qN+s\tau+t\right)^2}
$$
is the Weierstrass $\wp$-function with characteristics and the sum in (\ref{orbi-t3}) exhibits the properties
of the sum over conformal blocks of the CFT with the twisted boundary conditions
corresponding to the ${\mathbb Z}_N$-orbifold case.

\newsection{Discussion. Massive modes}

We have demonstrated that correlation functions in the AdS space in the sum over geometries
exhibit properties of sums over conformal blocks of the underlying CFT. Although we have considered only the
massless case in details, it seems plausible that the same procedure (with slight modifications) can
be applied to the whole spectrum of masses appearing in the AdS/CFT correspondence pattern. The generalization
seems to be rather straightforward. The factorization property must be nevertheless corrected; to see this, let us
consider an example of the correlation function for fields at the second mass level. Using (\ref{trans-Green}) and the
formula for $1/\sin^4(\alpha z)$,
$$
\frac1{\sin^4(\alpha z)}=\sum_{n=-\infty}^{\infty}\left(\frac1{(\alpha z+\pi n)^4}+\frac{2}{3(\alpha z+\pi n)^2}\right),
$$
and performing the summation over $(c,d)$-pairs, we obtain
\bea
&{}&\sum_{(c,d)}{}^{\prime\,\mathrm{(reg)}}G_{\mathrm{torus}}^{(c,d)}(z|m)|_{\kappa=4}=
\nonumber
\\
&=&F(z|2)+\sum_{(c,d)}{}'\left(G_{\mathrm{torus}}^{(c,d)}(z|m)|_{\kappa=4}-F(z|2)\right)
\nonumber
\\
&=&F(z|2)+
\nonumber
\\
&+&\sum_{(c,d)}{}'\left[\sum_{k_1,k_2,n\in {\mathbb Z}}
\left(\frac{1}{(z+k_1(c\tau+d)+n(a\tau+b))^4}+\frac{2\pi^2}{3(c\tau+d)^2(z+k_1(c\tau+d)+n(a\tau+b))^2}\right)\times\right.
\nonumber
\\
&\times&\left.
\left(\frac{1}{(\ov z+k_2(c\ov\tau+d)+n(a\ov\tau+b))^4}+\frac{2\pi^2}{3(c\ov\tau+d)^2
(\ov z+k_2(c\ov\tau+d)+n(a\ov\tau+b))^2}\right)-F(z|2)\right].
\label{sum-4}
\eea
Besides the ``factored'' term $\frac16\wp''(z|\tau)\ov{\wp''(z|\tau)}$
appearing when summing over $(c,d)$-pairs the products of leading (first) terms in parentheses in (\ref{sum-4})
with the singularity cancelled by the term $F(z|2)$, we have the contributions
$$
\sum_{(c,d)}{}'\sum_{k_1,k_2,n\in {\mathbb Z}}
\frac{2\pi^2}{3(c\tau+d)^2(z+k_1(c\tau+d)+n(a\tau+b))^2(\ov z+k_2(c\ov\tau+d)+n(a\ov\tau+b))^4}+\hbox{c.c.}
$$
and
$$
\sum_{(c,d)}{}'\sum_{k_1,k_2,n\in {\mathbb Z}}
\frac{4\pi^4}{9|c\tau+d|^4(z+k_1(c\tau+d)+n(a\tau+b))^2(\ov z+k_2(c\ov\tau+d)+n(a\ov\tau+b))^2},
$$
which converge due to the presence of the $(c\tau+d)$ factors in denominators. But, again, just because of these factors,
we cannot perform the resummation procedure as in Secs.~4 and~5, and the possible modular behavior of these sums
becomes more involved. For instance, we can perform the resummation in the first contribution passing to
the summation variables $s_{1,2}$, $t_{1,2}$, which yields
\bea
&{}&\sum_{{{s_1,s_2\atop t_1,t_2}\in{\mathbb Z}\atop(\Delta s)^2+(\Delta t)^2>0}}
\frac{2\pi^2(\hbox{GCD}(\Delta s,\Delta t))^2}{3(\Delta s\tau+\Delta t)^2(z+s_1\tau+t_1)^2(\ov z+s_2\ov\tau+t_2)^4}
\nonumber
\\
&{}&\quad+\left(\sum_{(c,d)}{}'\frac1{(c\tau+d)^2}\right)
\sum_{s,t\in{\mathbb Z}}\frac{2\pi^2}{3(z+s\tau+t)^2(\ov z+s\ov\tau+t)^4}+\hbox{c.c.},
\nonumber
\eea
where the coefficient by the second term is
$$
\sum_{(c,d)}{}'\frac1{(c\tau+d)^2}=\frac1{\xi(2)}E_2(\tau)=\frac3{\pi^2}E_2(\tau),
$$
and $E_2(\tau)\equiv\sum_{m,n\in{\mathbb Z},\,m^2+n^2>0}\frac1{(m\tau+n)^2}$ is the Eisenstein series.

Even more difficult (in the ideological sense)
problem is the problem of determining the proper summation measure.
Under our bulk regularization for the Green's function, it seems plausible that in order to
reproduce the proper scaling behavior of the Green's functions for different geometries,
this measure (or, the corresponding hyperbolic volume)
must be constant for all solid torus geometries
with the given boundary surface metric; this results however in the necessity
to introduce regularizing factors; nevertheless, it turns out that
these regularizing factors are independent on the additional
structure (on the choice of the $a$-cycles) and can be therefore determined unambiguously as soon as we fix
the two-dimensional metric. Still, it is an important question whether it is possible to obtain the
relevant regularizing factors from the field-theory considerations related to D1--D5 brane systems
(see~\cite{Moore} and references therein).
It would be interesting to check whether the Hamiltonian prescription of~\cite{ArFr} concerning local
contributions may help in constructing such a regularization. We however hope that the very appearance of the
conformal block structure in the sum over geometries in our approach justifies
further studies of these, relatively simple, systems. Another rather straightforward generalization can be
considering the supersymmetrization of the whole pattern.

Another, really challenging, problem is to consider generalizations of this technique to handlebodies of higher genus.
There, we must use the Schottky uniformization picture on the complex $\xi$-plane while the regularizing surface
must be determined by the equation
$$
t=\eps\e^{\phi(\xi,\ov\xi)}
$$
with $\phi(\xi,\ov\xi)$ satisfying the Liouville equation. That is, we must be able to work with expressions
of the form (cf.~(\ref{asympt}))
$$
G_{\mathrm{higher\ genus}}(\xi,\chi|m)\simeq
\sum_{k}\left(\frac{|\xi-\chi_k|^2}{2\eps^2|\e^{\phi(\xi,\ov\xi)}|\,|\e^{\phi(\chi_k,\ov\chi_k)}|}\right)^{-\kappa},
$$
where the sum ranges all images of the point $\chi$ under the Schottky group action. And it then still remains
the problem of performing the
summation over all AdS$_3$ geometries determined by all possible choices of the $a$-cycle
structures on the relevant Riemann surface. This may lead to a progress in studying Liouville systems as well.

\newsection{Acknowledgments}
The author acknowledges numerous useful discussions with K.~Krasnov on the subject of this paper.
The work was supported by the Russian Foundation for Basic Research (Grant
No.\,03-02-17373), by the Program of Supporting Leading Scientific Schools (Grant No.~N.Sh.-2052.2003.1)
and by the Program Mathematical Methods in Nonlinear Dynamics.


\begin{thebibliography}{99}
\bibitem{Maldacena} J.~M.~Maldacena, {\sl Adv. Theor. Math. Phys.}
{\bf 2} (1998) 231--252 (hep-th/9711200).
\bibitem{Polyakov} S.~S.~Gubser, I.~R.~Klebanov, and A.~M.~Polyakov,
{\sl Phys. Lett.} {\bf B428} (1998) 105--114 (hep-th/9802109).
\bibitem{Witten} E.~Witten, {\sl Adv. Theor. Math. Phys.} {\bf 2} (1998)
253--291 (hep-th/9802150).
\bibitem{inter} W.~Mueck and K.~S.~Vishwanatan, {\sl Phys. Rev.} {\bf D58}
(1998) 41901 (hep-th/9804035);\\
Hong Liu and A.~A.~Tseytlin,
{\sl Phys. Rev.} {\bf D59} (1999) 086002 (hep-th/9807097);\\
D.~Z.~Freedman, S.~D.~Mathur, A.~Matusis, and L.~Rastelli,
{\sl Phys. Lett.} {\bf B452} (1999) 61--68 {hep-th/9808006};\\
E.~D'\,Hoker, D.~Z.~Freedman, S.~D.~Mathur, A.~Matusis, and L.~Rastelli,
{\sl Nucl. Phys.} {\bf B562} (1999) 353--394 (hep-th/9903196).
\bibitem{HF} E.~D'\,Hoker and D.~Z.~Freedman,
{\sl Nucl. Phys.} {\bf B544} (1999) 612--632 (hep-th/9809179).
\bibitem{Bonelli} G.~Bonelli,
{\sl Phys. Lett.} {\bf B450} (1999) 363--367 (hep-th/9810194).
\bibitem{Ch1} L.~Chekhov, ``{\it AdS/CFT correspondence on torus},''
hep-th/9811146.
\bibitem{Ch} L.~Chekhov,
{\sl Mod. Phys. Lett.} {\bf A14} (1999) 2157--2168 (hep-th/9903209).
\bibitem{Wit2} E.~Witten, {\sl Adv. Theor. Math. Phys.} {\bf2} (1998) 505--532
(hep-th/9803131).
\bibitem{HP83} S.~W.~Hawking and D.~Page, {\sl Commun. Math. Phys.}
{\bf 87} (1983) 577--588.
\bibitem{CT} S.~Carlip and C.~Teitelboim, {\sl Phys. Rev.} {\bf D51}
(1995) 622--632.
\bibitem{MS} J.~Maldacena and A.~Strominger, {\sl JHEP} {\bf 9812} (1998) 005
(hep-th/9804085).
\bibitem{Farey} R.~Dijkgraaf, J.~M.~Maldacena, G.~W.~Moore, and E.~Verlinde,
``{\it A black hole Farey tail},'' hep-th/0005003.
\bibitem{K-V} E.~Keski-Vakkuri, {\sl Phys. Rev.} {\bf D59} (1999) 104001.
\bibitem{Krasnov} K.~Krasnov,
{\sl Adv. Theor. Math. Phys.} {\bf 4} (2000) 929--979 (hep-th/0005106).
\bibitem{Manin} D.~Hejhal, {\sl Adv. Math.} {\bf 15} (1975) 133--156.
\bibitem{Sel} W.~P.~Thurston, ``The geometry and topology of
{\rm3}-manifolds,'' Princeton notes, 1979.
\bibitem{Kach} M.~Kleban, M.~Porrati, and R.~Rabad\'an, ``{\it Poincar\'e recurrences
and topological diversity\/},'' hep-th/0407192.
\bibitem{GL} C.~R.~Graham and J.~M.~Lee, {\sl Adv. Math.} {\bf 87} (1991) 186.
\bibitem{Moore} G.~W.~Moore, ``{\it Les Houches lectures on strings and arithmetics},''
hep-th/0401049.
\bibitem{ArFr} G.~E.~Arutyunov and S.~A.~Frolov, {\sl Nucl. Phys.} {\bf B544} (1999) 576--589
(hep-th/9806216).
\end{thebibliography}
\end{document}